\documentclass[twocolumn,english,prl,amsmath,amssymb,showpacs]{revtex4}
\usepackage[T1]{fontenc}
\usepackage[latin9]{inputenc}
\usepackage{amsmath}
\usepackage{amssymb}
\usepackage{graphicx}
\usepackage{esint}

\makeatletter
\@ifundefined{textcolor}{}
{%
 \definecolor{BLACK}{gray}{0}
 \definecolor{WHITE}{gray}{1}
 \definecolor{RED}{rgb}{1,0,0}
 \definecolor{GREEN}{rgb}{0,1,0}
 \definecolor{BLUE}{rgb}{0,0,1}
 \definecolor{CYAN}{cmyk}{1,0,0,0}
 \definecolor{MAGENTA}{cmyk}{0,1,0,0}
 \definecolor{YELLOW}{cmyk}{0,0,1,0}
}

\usepackage{babel}

\makeatother

\usepackage{babel}
\begin{document}

\title{Skyrmionic Spin Seebeck Effect via Dissipative Thermomagnonic Torques}

\author{Alexey A. Kovalev }

\affiliation{Department of Physics and Astronomy and Nebraska Center for Materials
and Nanoscience, University of Nebraska, Lincoln, Nebraska 68588,
USA}

\date{\today}
\begin{abstract}
We derive thermomagnonic torque and its ``$\beta$-type'' dissipative
correction from the stochastic Landau-Lifshitz-Gilbert equation. The ``$\beta$-type'' dissipative correction describes viscous coupling between magnetic dynamics and magnonic current and it stems
from spin mistracking of the magnetic order. We
show that thermomagnonic torque is important for describing temperature gradient
induced motion of skyrmions in helical magnets while dissipative
correction plays an essential role in generating transverse Magnus
force. We propose to detect such skyrmionic motion by employing the
transverse spin Seebeck effect geometry. 
\end{abstract}


\maketitle
\textit{Introduction.} Out-of-equilibrium effects involving spin are
essential to understanding many spintronic discoveries, such as spin-transfer
torque \cite{Slonczewski:JMMM1996,Berger:Oct1996,Tsoi.Jansen.ea:PRL1998,Myers:S1999},
spin pumping \cite{Tserkovnyak:PRL2002,Heinrich:May2003} and the
spin Seebeck effect \cite{Uchida.Takahashi.ea:2008,Uchida.Ota.ea:JoAP2010,Jaworski.Yang.ea:NM2010,Uchida:nov2010,Bosu.Sakuraba.ea:2011}.
These discoveries enabled unprecedented degree of control in magnetic
information-storage devices in which the magnetization can be flipped
at will \cite{Maekawa.Adachi.ea:JPSJ2013} or domain wall can be moved
in order to change the magnetization configuration \cite{Parkin.Hayashi.ea:S2008}.
Sizable coupling of spin to thermal flows \cite{Bauer.Saitoh.ea:NM2012}
leads to yet another knob by which we can control magnetization and
magnetic textures such as domain walls \cite{Kovalev:PRB2009,Bauer:jan2010,Hinzke:PRL2011,Kovalev:EPL2012}
as confirmed in recent experiments \cite{Torrejon.Malinowski.ea:PRL2012,Jiang:PRL2013}.
In addition, such coupling can enable energy harvesting from temperature
gradients by employing the spin Seebeck effect \cite{Cahaya.Tretiakov.ea:APL2014}
or magnetic texture dynamics \cite{Kovalev:SSC2010}. 

Particularly strong coupling between heat flows carried by magnons
and magnetic textures are expected in magnetic insulators such as
$\text{Cu}_{2}\text{O}\text{Se}\text{O}_{3}$ \cite{Seki:Science2012},
$\text{Ba}\text{Fe}_{1-x-0.05}\text{Sc}_{x}\text{Mg}_{0.05}\text{O}_{19}$
\cite{Yu.Mostovoy.ea:PotNAoS2012} and $\text{Y}_{3}\text{Fe}_{5}\text{O}_{12}$
\cite{Bhagat.Lessoff.ea:PSSa1973} where the magnetization dynamics
have low dissipation as coupling to electron continuum is absent  \cite{Kovalev:EPL2012,Hoffman.Sato.ea:PRB2013}.
At the same time even at relatively low temperatures thermal magnons
have very small wavelength and thus can be treated as particles on
the scale of magnetic texture \cite{Dugaev:jul2005,Kovalev:EPL2012}.
This brings a lot of analogies to magnetization dynamics in metallic
systems where the dynamics can be controlled by flows of electrons
\cite{Zhang.Li:PRL2004,Kohno.Tatara.ea:JotPSoJ2006,Tserkovnyak.Skadsem.ea:PRB2006,Tserkovnyak.Brataas.ea:JoMaMM2008}.
In conducting materials, on the other hand, thermoelectric spin torque
can also couple magnetization to heat flows even in the absence of
charge flows \cite{Kovalev:PRB2009,Kovalev:SSC2010,Hals.Brataas.ea:SSC2010}.

A skyrmion is an example of a magnetic texture that can arise in helical
magnets due to inversion asymmetry induced Dzyaloshinsky - Moriya
(DM) interaction \cite{Rossler:2006} as observed in bulk samples
of $\mbox{MnSi}$ by neutron scattering techniques \cite{Muhlbauer:Science2009}.
Skyrmion crystal (SkX) is particularly stable in two-dimensional (2D)
systems or thin films as was predicted theoretically \cite{Yi:2009,Butenko:2010}
and later confirmed experimentally \cite{Yu:Nature2010,Yu:NM2011}.
Just like other textures, such as domain walls, skyrmions can be manipulated
by temperature gradients \cite{Mochizuki.Yu.ea:NM2014,Everschor2012}. However,
many aspects related to interaction of magnon spin currents to magnetic
textures are still unclear \cite{Kong:PRL2013,Saxena:arXiv2013}. 
\begin{figure}
\centerline{\includegraphics[clip,width=1\columnwidth]{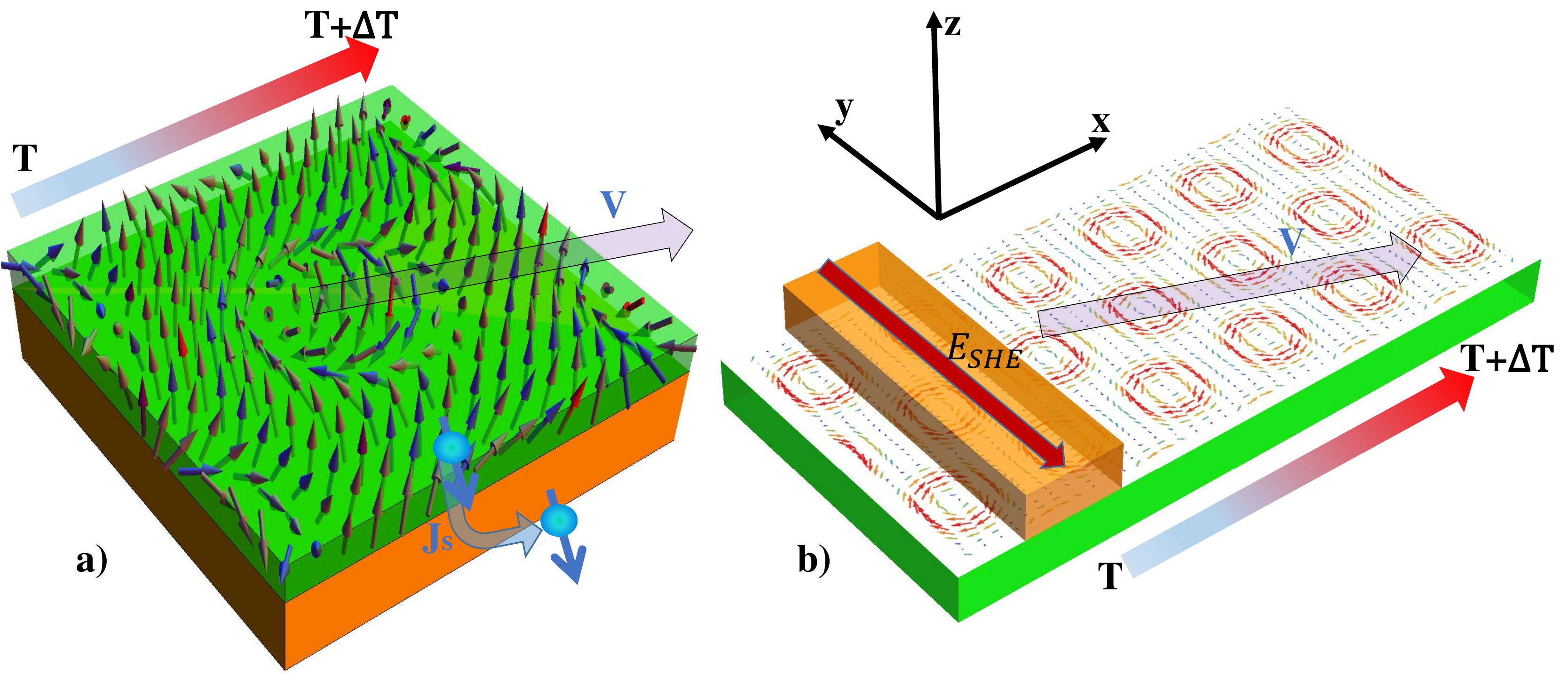}}

\caption{(Color online) a) The magnon current induced by temperature gradient
exerts spin torque on magnetization which leads to skyrmion motion
in the direction of the hot region with an additional Hall-like side
motion. An additional non-magnetic layer, such as Pt, can be used
in order to detect the spin
pumping resulting from skyrmionic motion, i.e. via the inverse spin Hall effect. b) Spin Seebeck effect geometry
can be used for detection of the Hall-like motion of skyrmions. Due
to mostly out-of-plane magnetization configuration the ordinary spin
Seebeck effect should be suppressed. }

\label{fig:Fig1} 
\end{figure}

In this Rapid Communication, we address the dissipative ``$\beta$-type''
corrections to magnonic spin torques. Such corrections play an important
role in domain wall dynamics \cite{Zhang.Li:PRL2004,Kohno.Tatara.ea:JotPSoJ2006,Tserkovnyak.Skadsem.ea:PRB2006,Tserkovnyak.Brataas.ea:JoMaMM2008};
however, they were introduced only phenomenologically in relation
to magnonic torques \cite{Kovalev:EPL2012,Saxena:arXiv2013}. Here we
derive thermomagnonic torque and its ``$\beta$-type'' correction
from the stochastic Landau-Lifshitz-Gilbert (LLG) equation. By applying
this theory to skyrmionic textures, we conclude that ``$\beta$-type''
correction influences the sign of the transverse Magnus force which
can be tested in the spin Seebeck effect geometry termed as skyrmionic
spin Seebeck effect (see Fig.~\ref{fig:Fig1}). The
sign of the Magnus force is indicative of two regimes (i) $\beta>\alpha$
and (ii) $\beta<\alpha$ where according to our calculations the stochastic
LLG equation results in the regime (i), here $\alpha$ is the Gilbert damping. We also analyze thermoelectric
spin torque contributions that can arise in conducting ferromagnets
even in the absence of the charge transport. We compare thermoelectric
contributions to their thermomagnonic counterparts. 

\textit{Thermal magnons and magnetic texture.} We consider a ferromagnet
well below the Curie temperature in which the magnetization dynamics
is described by the stochastic LLG equation: 
\begin{equation}
s(1+\alpha\mathbf{m}\times)\mathbf{\dot{m}}+\mathbf{m}\times(\mathbf{H}_{\text{eff}}+\mathbf{h})=0\:,\label{eq:LLG}
\end{equation}
where $s$ is the saturation spin density, $\mathbf{m}(\mathbf{r},t)$
is a unit vector in the direction of the spin density and $\mathbf{H}_{\text{eff}}=-\delta_{\mathbf{m}}F$
describes the effective magnetic field. For the purposes of this paper,
we consider the free energy density $F=(A/2)(\partial_{\alpha}\mathbf{m})^{2}+D\mathbf{m}\cdot(\nabla\times\mathbf{m})-\mathbf{m}\cdot\mathbf{H}$
where $A_{x}=A/M_{s}$ is the exchange stiffness, $D_{\text{dm}}$
describes DM interaction with $D\equiv D_{\text{dm}}M_{s}$, $M_{s}$
is the saturation magnetization, and $\mathbf{H}_{e}$ is the external
magnetic field with $\mathbf{H}\equiv\mathbf{H}_{e}M_{s}$. Note,
however, that the discussion in this section is general and should
also apply to the most general form of the free energy density as
long as other energy contributions can be ignored in comparison to
exchange energy of thermal magnons which is true at sufficiently high
temperatures and for sufficiently small strength of DM interactions.
In Eq.~(\ref{eq:LLG}) $\mathbf{h}$ is the random Langevin field
corresponding to thermal fluctuations at temperature $T$ with correlator
\cite{Brown:PR1963}:
\begin{equation}
\left\langle h_{i}(\mathbf{r},t)h_{j}(\mathbf{r}',t')\right\rangle =2\alpha sk_{B}T(\mathbf{r})\delta_{ij}\delta(\mathbf{r}-\mathbf{r}')\delta(t-t'),\label{eq:stoch}
\end{equation}
which is consistent with the fluctuation dissipation theorem. The
fluctuating field results in fast magnetization dynamics that happens
on top of the slow magnetic texture dynamics with long characteristic
length-scale. The purpose of this paper is to describe how the fast
magnetization dynamics influence the slow dynamics. In Eq.~(\ref{eq:stoch})
we assume a uniform temperature gradient along the $x-$axis, i.e.
$\partial T(x)/\partial x=\mbox{constant}$.

We consider small fast oscillations of the magnetization with time scale $1/\omega_q$ defined by the magnon frequency $\omega_q$
on top of the slow magnetization dynamics with time scale defined by external magnetic field and current. The vectors
for the fast $\mathbf{m}_{f}(\mathbf{r},t)$ and slow $\mathbf{m}_{s}(\mathbf{r},t)$
magnetization dynamics are related by $\mathbf{m}=(1-\mathbf{m}_{f}^{2})^{1/2}\mathbf{m}_{s}+\mathbf{m}_{f}$
where $\mathbf{m}_{s}\cdot\mathbf{m}_{f}=0$. We
apply a coordinate transformation after which the $z-$axis points
along the spin density of the slow dynamics. In the new coordinate
system, small excitations will only have $m_{x}$ and $m_{y}$ components.
In order to describe our system in the new coordinates, we introduce
$3\times3$ rotation matrix $\hat{R}=\exp\theta\hat{J}_{y}\exp\phi\hat{J}_{z}$
with $\hat{J}_{\alpha}$ being the $3\times3$ matrix describing infinitesimal
rotation along the axis with index $\alpha$. In the new coordinates,
we have $\mathbf{m}\rightarrow\mathbf{m}^{'}=\hat{R}\mathbf{m}$ and
the covariant derivative $\partial_{\mu}\rightarrow(\partial_{\mu}-\hat{\mathcal{A}}_{\mu})$
with $\hat{\mathcal{A}}_{\mu}=(\partial_{\mu}\hat{R})\hat{R}^{-1}$
(the index $\mu=0,..,3$ denotes the time and space coordinates).
Since the matrix $\hat{\mathcal{A}}_{\mu}$ is skew-symmetric, we
can introduce a vector $\mathbf{\boldsymbol{\mathcal{A}}}_{\mu}$
so that $\hat{\mathcal{A}}_{\mu}\mathbf{m}=\mathbf{\boldsymbol{\mathcal{A}}}_{\mu}\times\mathbf{m}$.
In some specific gauge, the elements of $\mathbf{\boldsymbol{\mathcal{A}}}_{\mu}$
become $\mathbf{\boldsymbol{\mathcal{A}}}_{\mu}=(-\sin\theta\partial_{\mu}\phi,\partial_{\mu}\theta,\cos\theta\partial_{\mu}\phi)$.
The equation describing fast dynamics follows from the LLG equation
subject to the coordinate transformation: 
\begin{equation}
is[\partial_{t}(1-i\alpha)-i\mathcal{A}_{0}^{z}]m_{+}=A\left(\partial_{\alpha}/i-\mathcal{A}_{\alpha}^{z}\right)^{2}m_{+}+Hm_{+}\:,\label{eq:Magnons}
\end{equation}
where we disregarded various anisotropy terms assuming that the exchange
interactions are dominant. As a result, the coupling between the circular
components \cite{Dugaev:jul2005} of spin wave $m_{\pm}=m_{x}^{'}(\mathbf{r},t)\pm im_{y}^{'}(\mathbf{r},t)$
can be disregarded where $m_{x(y)}^{'}(\mathbf{r},t)$ are the transverse
excitations in the transformed coordinates with the $z$ axis pointing
along the direction of $\mathbf{m}_{s}$. Equation (\ref{eq:Magnons})
describes thermal magnons with spectrum $\omega_{q}=(H+Aq^{2})/s$
where the magnon current can be written as $j_{\alpha}=\frac{A}{2i}(m_{-}\partial_{\alpha}m_{+}-m_{+}\partial_{\alpha}m_{-})$.
Note that formally Eq.~(\ref{eq:Magnons}) describes charged particles
moving in the fictitious electric $\mathcal{E}_{\alpha}=-\partial_{t}\mathcal{A}_{\alpha}^{z}-\partial_{\alpha}\mathcal{A}_{0}^{z}=\hbar\mathbf{\widetilde{m}}_{s}\cdot(\partial_{t}\mathbf{\widetilde{m}}_{s}\times\partial_{\alpha}\mathbf{\widetilde{m}}_{s})$
and magnetic $\mathcal{B}_{i}=(\hbar/2)\epsilon^{ijk}\mathbf{\widetilde{m}}_{s}\cdot(\partial_{k}\mathbf{\widetilde{m}}_{s}\times\partial_{j}\mathbf{\widetilde{m}}_{s})$
fields produced by the magnetic texture. 

We are now in the position to calculate the force that fast oscillations
exert on the slow magnetization dynamics $\mathbf{m}_{s}(\mathbf{r},t)$.
For simplicity, we assume that the texture is static as we can add
the time dependent contribution, e.g. corresponding to $\mathcal{A}_{0}^{z}$,
later by involving the Onsager reciprocity principle. The force due
to rapid oscillations of the fast component averages out in the first
order terms with respect to $\mathbf{m}_{f}(\mathbf{r},t)$, hence
such force can only come from the second order terms. In the effective
field $\mathbf{H}_{\text{eff}}=\mathbf{H}+A\boldsymbol{\nabla}^{2}\mathbf{m}-2D\boldsymbol{\nabla}\times\mathbf{m}$
relevant terms can be immediately identified leading to the following
contribution:
\begin{align}
\boldsymbol{\mathcal{T}}=&\mathbf{m}_{s}\times\mathbf{H}_{\text{eff}}^{s}-\bigl\langle\mathbf{m}\times\mathbf{H}_{\text{eff}}\bigr\rangle\nonumber \\ \nonumber
=&A\mathbf{m}_{s}\times\boldsymbol{\mathcal{S}}\approx A\bigl\langle\mathbf{m}_{f}\times\boldsymbol{\nabla}^{2}\mathbf{m}_{f}\bigr\rangle\\ 
+&2A\bigl\langle\mathbf{m}\times  \partial_\alpha\mathbf{m}_s \partial_\alpha(\mathbf{m}_s\cdot\mathbf{m})\bigr\rangle,\label{eq:torque}
\end{align}
where $\mathbf{H}_{\text{eff}}^{s}=-\delta_{\mathbf{m}_{s}}F(\langle m \rangle\mathbf{m}_{s},\langle m \rangle\partial_{\alpha}\mathbf{m}_{s})$, we formally defined $\boldsymbol{\mathcal{S}}$ as the non-equilibrium
transverse accumulation of magnon spins, $\bigl\langle\ldots\bigr\rangle$
stands for averaging over the fast oscillations induced by random
fields, and the terms corresponding to DM interactions have been dropped
since they contain one spacial derivative which for thermal fluctuations
with small wavelength leads to a small parameter $D/(Aq)$. We now
calculate the torque in Eq.~(\ref{eq:torque}) in the reference frame
associated with the slow dynamics by employing the covariant derivative.
By dropping the terms that average out due to fast oscillations and
concentrating only on the torque components that are in the $x'-y'$
plane we obtain the following result for the transverse accumulation
of magnon spins: 
\begin{equation}
\boldsymbol{\mathcal{S}}=2\bigl\langle\mathbf{m}_{f}(\partial_{\alpha}\mathbf{m}_{s}\cdot\partial_{\alpha}\mathbf{m}_{f})\bigr\rangle-2\bigl\langle\partial_\alpha\mathbf{m}_s(\mathbf{m}_f\cdot\partial_\alpha\mathbf{m}_f)\bigr\rangle.\label{eq:spin-accumulation}
\end{equation}
In the transformed reference frame vectors $\boldsymbol{\mathcal{S}}$,
$\mathbf{m}_{f}$, $\partial_{\alpha}\mathbf{m}_{s}$ and $\partial_{\alpha}\mathbf{m}_{f}$
are in the $x'-y'$ plane, thus it is convenient to switch to complex
notations $a\equiv a_{x}^{'}+ia_{y}^{'}$ where $\mathbf{a}$ is an
arbitrary vector in the $x'-y'$ plane which leads to $S=\bigl\langle m_{f}(\partial_{\alpha}m_{s}\partial_{\alpha}m_{f}^{*}+\partial_{\alpha}m_{s}^{*}\partial_{\alpha}m_{f})\bigr\rangle-\bigl\langle \partial_{\alpha}m_{s}(m_{f}\partial_{\alpha}m_{f}^{*}+m_{f}^{*}\partial_{\alpha}m_{f})\bigr\rangle=-\partial_{\alpha}m_{s}\bigl\langle m_{-}\partial_{\alpha}m_{+}\bigr\rangle$.
For the steady state solution we find:
\begin{equation}
\mathcal{S}=-\partial_{x}m_{s}\int\dfrac{d^{d-1}\mathbf{q}d\omega}{(2\pi)^{d}}\dfrac{\left\langle m_{-}(\mathbf{q},\omega,x)\partial_{x}m_{+}(\mathbf{q}',\omega',x)\right\rangle }{(2\pi)^{d}\delta(\mathbf{q}-\mathbf{q}')\delta(\omega-\omega')},\label{eq:spin-accumulation1}
\end{equation}
where $d=2$ or $3$ depending on the dimensionality of the magnet
and $S=S_{x}+iS_{y}$ describes two components of spin accumulation
leading to the dissipative and nondissipative torques. Here we used
$m_{\pm}(\mathbf{r},t)$ Fourier transformed with respect to time
and transverse coordinate: 
\begin{equation}
m_{\mp}(\mathbf{q},\omega,x)=\int\dfrac{d^{d-1}\boldsymbol{\rho}d\omega}{(2\pi)^{d}}e^{\pm i(\omega t-\mathbf{q}\boldsymbol{\rho})}m_{\mp}(\mathbf{r},t).
\end{equation}
The delta functions in denominator of Eq.~(\ref{eq:spin-accumulation1})
cancel out after the averages over stochastic fields are evaluated.
For the corresponding stochastic fields we obtain \cite{Hoffman.Sato.ea:PRB2013}:
\begin{equation}
\dfrac{\left\langle h(x,\mathbf{q},\omega)^{*}h(x',\mathbf{q}',\omega')\right\rangle }{4(2\pi)^{d}\alpha sk_{B}}=T(x)\delta(x-x')\delta(\mathbf{q}-\mathbf{q}')\delta(\omega-\omega').\label{eq:stoch-1}
\end{equation}
Since we are after the first order terms in magnetic texture gradients
we can use Eq.~(\ref{eq:Magnons}) in the absence
of vector potentials. The stochastic LLG Eq.~(\ref{eq:LLG}) takes
the form of the homogeneous Helmholtz equation:
\begin{equation}
A(\partial_{x}^{2}+k^{2})m_{-}(x,\mathbf{q},\omega)=h(x,\mathbf{q},\omega),\label{eq:Helmholtz}
\end{equation}
where this equation corresponds to Eq.~(\ref{eq:Magnons})
with an added stochastic term and $k^{2}=[(1+i\alpha)s\omega-H]/A-q^{2}$.
Equation (\ref{eq:Helmholtz}) can be easily solved by employing Green's
function $G(x-x_{0})=ie^{ik|x-x_{0}|}/(2k)$. We substitute this solution
in Eq.~(\ref{eq:spin-accumulation1}) and carry through integrations
over variables $x$ and $x'$ in Eq.~(\ref{eq:stoch-1}). The final
expression for the magnon spin torque becomes:
\begin{equation}
\mathcal{T}=-\dfrac{s\alpha}{4A}\partial_{x}m_{s}\int\dfrac{d^{d-1}\mathbf{q}}{(2\pi)^{d}}{\displaystyle \int_{\omega_{0}}^{\infty}}d\omega\dfrac{\partial_{x}\bigl(\hbar\omega\coth\tfrac{\hbar\omega}{2k_{B}T}\bigr)}{k^{*}(\text{Im}k)^{2}}.\label{eq:spin-accumulation2}
\end{equation}
Here we limited frequency integration by $\omega_{0}=(H+Aq^{2})/s$
and replaced $2k_{B}T\rightarrow\hbar\omega\coth(\hbar\omega/2k_{B}T)$
by employing the quantum fluctuation dissipation theorem which introduces
high frequency cut off at $\hbar\omega\gg k_{B}T$. The former allows
us to limit our consideration to magnonic excitations with energies
above the magnonic gap and the latter allows us to relate our expression
to magnon currents obtained by the Boltzmann approach. By replacing
integration over $\omega$ with integration over $k$ and keeping
only the first two orders in $\alpha$ we obtain:

\begin{equation}
\mathcal{T}=-\hbar\partial_{x}m_{s}j_{x}(1+i\beta)\:,\label{eq:Boltzmann}
\end{equation}
where $j_{x}=(\partial_{x}T/T)\int d^{d}\mathbf{k}/(2\pi)^{d}\tau(\varepsilon)\varepsilon\upsilon_{x}^{2}\partial f_{0}/\partial\varepsilon$
with $\tau(\varepsilon)=(2\alpha\omega)^{-1}$ can be interpreted
as the magnon current calculated within the relaxation time approximation
from non-equilibrium distribution correction $\delta f=\tau\varepsilon(\partial f/\partial\varepsilon)\upsilon_{\alpha}(\partial_{\alpha}T/T)$
\cite{Ashcroft.Mermin:1976}. Here $\varepsilon(\mathbf{q})=\hbar(Ak^{2}+H)/s$,
$\upsilon_{x}=\partial\omega_{q}/\partial k_{x}$, and $f_{0}=\left\{ \exp\left[\varepsilon/k_{B}T\right]-1\right\} ^{-1}$
is the Bose-Einstein equilibrium distribution. This result is expected
given that each magnon carries angular momentum $-\hbar$. The second
term in Eq.~(\ref{eq:Boltzmann}) corresponds to the dissipative correction
with $\beta/\alpha=(d/2)F_{1}(x)/F_{0}(x)\sim d/2$ with $F_{0}(x)=\int_{0}^{\infty}d\epsilon\epsilon^{d/2-1}\epsilon e^{\epsilon+x}/(e^{\epsilon+x}-1)^{2}$
and $F_{1}(x)=\int_{0}^{\infty}d\epsilon(\epsilon+x)\epsilon^{d/2-1}e^{\epsilon+x}/(e^{\epsilon+x}-1)^{2}$
evaluated at the magnon gap $x=\hbar\omega_{0}/k_{B}T$ where $d=2$
or $3$. The magnon current density is given by \cite{Kovalev:EPL2012}:
\begin{equation}
j_{\alpha}=k_{B}\partial_{\alpha}TF_{0}/(6\pi^{2}\lambda\hbar\alpha),\label{eq:MagnonCurrent-1}
\end{equation}
 where $d=3$ and $\lambda=\sqrt{\hbar A/(sk_{B}T)}$ is the thermal
magnon wavelength {[}for $d=2$ we obtain $j_{\alpha}=k_{B}\partial_{\alpha}TF_{0}/(4\pi\hbar\alpha)${]}.
We can express result in Eq.~(\ref{eq:Boltzmann}) in the form of
LLG equation:
\begin{equation}
\begin{array}{cl}
\mathfrak{s}(1+\alpha^{s}\mathbf{m}_{s}\times)\mathbf{\dot{m}}_{s} =&\mathbf{H}_{\text{eff}}^{s}\times\mathbf{m}_{s}-\left[1+\beta\mathbf{m}_{s}\times\right](\mathbf{j}_{m}^{s}\boldsymbol{\partial})\mathbf{m}_{s}\\
\\
 & -\left[1+\beta_{e}\mathbf{m}_{s}\times\right](\mathbf{j}_{e}^{s}\boldsymbol{\partial})\mathbf{m}_{s},
\end{array}\label{LLG-1}
\end{equation}
where $\mathfrak{s}=\bigl\langle\mathbf{m}\bigr\rangle s$ is the
renormalized spin density, $\mathbf{H}_{\text{eff}}^{s}=-\delta_{\mathbf{m}_{s}}F(\langle m \rangle\mathbf{m}_{s},\langle m \rangle\partial_{\alpha}\mathbf{m}_{s})$
is the effective field, $\alpha^{s}=\bigl\langle\mathbf{m}\bigr\rangle\alpha$
is the renormalized Gilbert damping, $\mathbf{j}_{m}^{s}=-\hbar\mathbf{j}$
is the spin current with polarization along $\mathbf{m}_{s}$ carried
by magnons and we added the terms proportional to thermoelectric spin
current $\mathbf{j}_{e}^{s}=\boldsymbol{\partial}T\wp_{S}S\sigma_{F}(1-\wp^{2})\hbar/2e$
arising under the open circuit conditions in conducting materials
(i.e. in the absence of charge currents) \cite{Kovalev:PRB2009} with
$S$ being the Seebeck coefficient, $\wp_{S}$ being the polarization
of the Seebeck coefficient, $\sigma_{F}$ being the conductivity of
the ferromagnet with polarization $\wp$ and $\beta_{e}$ being the
``$\beta$-type'' correction corresponding to thermoelectric spin
currents. 

\textit{Application to skyrmion dynamics.} Equation (\ref{LLG-1})
can be used in order to describe the magnetization dynamics in response
to temperature gradients. Here we analyze motion of skyrmions in response
to temperature gradients in ferromagnetic insulators. For describing
motion of magnetic textures such as magnetic domain walls and vortices
one can use the approach proposed by Thiele \cite{Thiele:PRL1973}.
Within such an approach, the dynamics are constrained to a subspace
described by the generalized coordinates $q$, $\dot{\mathbf{m}}=\sum_{i}\dot{q}_{i}\partial_{q_{i}}\mathbf{m}$
where the equation of motion for $q$ can be found by integrating
$\int d^{3}\mathbf{r}\partial_{q_{i}}\mathbf{m}\cdot(\mathbf{m}\times\dots)$
where instead of dots one needs to substitute Eq.~(\ref{LLG-1}).
One can apply this approach in order to describe motion of a skyrmion
in response to magnon currents in Eq.~(\ref{LLG-1}) as long as the
thermal magnon wavelength is smaller than the typical size of a skyrmion.
The following equation describing skyrmion dynamics in a thin film
results from Eq.~(\ref{LLG-1}): 
\begin{equation}
\boldsymbol{{\cal G}}\times(\mathbf{j}_{m}^{s}+\mathbf{j}_{e}^{s}-\mathfrak{s}\mathbf{v})+\eta(\beta\mathbf{j}_{m}^{s}+\beta_{e}\mathbf{j}_{e}^{s}-\alpha\mathfrak{s}\mathbf{v})=0,\label{eq:skyrmion}
\end{equation}
 where $\boldsymbol{{\cal G}}=\hat{z}\int d^{2}\mathbf{r}(\partial_{x}\mathbf{m}_{s}\times\partial_{y}\mathbf{m}_{s})\cdot\mathbf{m}_s$, $\eta=\eta_{\alpha}=\int d^{2}\mathbf{r}(\partial_{\alpha}\mathbf{m}_{s})^{2}/(4\pi)$
is the form factor of skyrmion which is of the order of $1$ and $\mathbf{v}$
is the skyrmion velocity. The skyrmion will move along the temperature
gradient with velocity: 
\begin{equation}
v_{x}=\dfrac{(j_{m}^{s}+j_{e}^{s}){\cal G}^2+\alpha\eta^{2}(\beta j_{m}^{s}+\beta_{e}j_{e}^{s})}{\mathfrak{s}({\cal G}^2+\alpha^{2}\eta^{2})},\label{eq:velocityX}
\end{equation}
in addition acquiring extra transverse Hall-like motion:
\begin{equation}
v_{y}=\eta{\cal G}\dfrac{\alpha(j_{m}^{s}+j_{e}^{s})-(\beta j_{m}^{s}+\beta_{e}j_{e}^{s})}{\mathfrak{s}({\cal G}^2+\alpha^{2}\eta^{2})}.\label{eq:velocityY}
\end{equation}
Here we define the Hall angle as $\tan\theta_{H}=v_{y}/v_{x}$.
For a ferromagnetic insulator ($j_{e}^{s}=0$) it is
easy to see  that the skyrmion will move
towards the hot region as follows from Eq.~(\ref{eq:skyrmion}). Earlier approaches either did not consider
the $\beta-$type corrections \cite{Kong:PRL2013} or introduced them
heuristically \cite{Saxena:arXiv2013}.
Note that Eq.~(\ref{eq:spin-accumulation2}) results in $\beta\approx1.5\alpha>0$.

\textit{Transverse spin Seebeck effect.} In ferromagnetic insulators
the coupling to the electron continuum is absent which may complicate
the detection of skyrmion dynamics. Here, we propose to detect temperature
induced skyrmion dynamics by employing spin pumping into the neighboring
non-magnetic metallic layer, such as Pt. The whole setup in Fig.~\ref{fig:Fig1}
then corresponds to the geometry of the transverse spin Seebeck effect
\cite{Uchida.Takahashi.ea:2008,Uchida.Ota.ea:JoAP2010,Jaworski.Yang.ea:NM2010,Uchida:nov2010,Bosu.Sakuraba.ea:2011}
in which the ordinary spin Seebeck response should be weak since we
are not considering the in-plane magnetization configuration. We assume that the spin injection is locally homogeneous
which is justified when $\lambda_{\text{sd}}\ll\xi$ where $\xi=2\pi A/D$
is the characteristic length of a skyrmion estimated by the spiral
period. To calculate the spin accumulation and the spin current
in normal metal we employ the spin diffusion equation for the spin
accumulation $\nabla^{2}\boldsymbol{\mu}_{s}=\boldsymbol{\mu}_{s}/\lambda_{\text{sd}}$
where $\lambda_{\text{sd}}$ is the spin-diffusion length in the normal
metal. The boundary conditions at the interface between the normal
metal and ferromagnetic insulator enforce continuity of the spin current
$\partial_{z}\boldsymbol{\mu}_{s}\mid_{z=0}=-(G_{0}/\sigma_{N})\mathbf{j}_{s,z}$
where $\sigma_{N}$ is the conductivity of the normal metal and $G_{0}=2e^{2}/h$
is the quantum conductance. The spin current $\mathbf{j}_{s,z}=\mathbf{j}_{s}^{\text{pmp}}+\mathbf{j}_{s}^{\text{bf}}$
is the sum of the spin-pumped and backflow contributions:
\begin{equation}
\mathbf{j}_{s,z}=\dfrac{\hbar g^{\uparrow\downarrow}}{4\pi}\mathbf{m}_{s}\times\dfrac{\partial\mathbf{m}_{s}}{\partial t}+\dfrac{g^{\uparrow\downarrow}}{4\pi}\boldsymbol{\mu}_{s}\mid_{z=0},\label{eq:pumping}
\end{equation}
where $g^{\uparrow\downarrow}$ is the spin mixing conductance per
unit of interface area and the conductance quantum in which the imaginary
part is disregarded which is usually justified for realistic interfaces
with metals \cite{Xia.Kelly.ea:2002}. 

After solving the diffusion equation, we obtain the full spin flow as well as the average voltage difference
due to the inverse spin Hall effect (ISHE) \cite{Saitoh:APL2006} by averaging the result over the magnetic texture
of a skyrmion. The voltage due to the ISHE becomes:
\begin{equation}
\dfrac{V_{\text{ISH}}}{W}=\dfrac{\left|\left\langle \mathbf{j}_{s}^\text{pmp}\right\rangle _{s}\right|\theta_{\text{SH}}}{d_{N}|e|}\dfrac{\cosh\tfrac{d_{N}}{\lambda_{\text{sd}}}-1}{\tfrac{\sigma_{N}}{G_{0}\lambda_{\text{sd}}}\sinh\tfrac{d_{N}}{\lambda_{\text{sd}}}+\tfrac{g^{\uparrow\downarrow}}{4\pi}\cosh\tfrac{d_{N}}{\lambda_{\text{sd}}}},\label{eq:ISHE}
\end{equation}
where $\theta_{\text{SH}}$ is the spin Hall angle of the normal metal,
$\left\langle \mathbf{j}_{s}^\text{pmp}\right\rangle _{s}$ is the
pumped spin current density averaged over one skyrmion and $W$ is
the width over which the voltage is measured. The averaged spin current
becomes: 
\begin{equation}
\left\langle \mathbf{j}_{s}^\text{pmp}\right\rangle _{s}=\dfrac{\hbar g^{\uparrow\downarrow}}{4\pi\xi}\boldsymbol{\chi}_{\alpha}v_{\alpha},\label{eq:pumping1}
\end{equation}
where we introduced another form factor of a skyrmion $\boldsymbol{\chi}_{\alpha}=\int d^{2}\mathbf{r}(\mathbf{m}_{s}\times\partial_{\alpha}\mathbf{m}_{s})/\xi$
which defines the polarization of the averaged spin current density
in Fig.~\ref{fig:Fig1} (in-plane and orthogonal to skyrmion velocity).
Thus the voltage in Eq.~(\ref{eq:ISHE}) is generated along the direction
of skyrmion flow which includes a conventional Seebeck like contribution
along the applied temperature gradient as well as transverse component
corresponding to Nernst effect. 

\textit{Estimates.} For estimates, we consider $\text{Cu}_{2}\text{OSeO}_{3}$
thin insulating layer of thickness $d_{F}=15\mbox{nm}$ in skyrmionic
phase in contact with a thin Pt layer, $d_{N}=1\mbox{nm}$, (see~Fig.
\ref{fig:Fig1}). By taking the lattice spacing $a=0.5\mbox{nm}$,
$\mathfrak{s}=0.5\hbar/a^{3}$, $A/(a^{2}k_{B})=50\mbox{K}$ and $D/(ak_{B})=3\mbox{K}$
we obtain the spiral period $\xi\approx50\mbox{nm}$ \cite{Seki:Science2012}.
Even at low temperatures comparable to $1K$ the magnon wavelength
satisfies the condition $\lambda\ll d_{F}$ which should justify three
dimensional treatment in Eq.~(\ref{eq:Boltzmann}). We further take
the temperature $T=50\mbox{K}$, $\alpha=0.01$ and a gradient $\partial_{\alpha}T=1\mbox{K}/\mu\mbox{m}$
arriving at the skyrmion longitudinal velocity towards the hot region
$v_{x}\approx0.1\mbox{m/s}$ from Eq.~(\ref{eq:skyrmion}). From Eq.~(\ref{eq:pumping1}) the skyrmion motion leads to the ISHE voltage
$V_{\text{ISH}}/W=2\times10^{-2}\mbox{V}/\mbox{m}$ for $\theta_{\text{SH}}=0.11$,
$\lambda_{\text{sd}}=1.5\mbox{nm}$, $\sigma_{N}=1\mu\Omega^{-1}\mbox{m}^{-1}$
and $g^{\uparrow\downarrow}=1.5\times 10^{19}\mbox{m}^{-2}$ \cite{Weiler:PRL2013}.
This signal is much smaller than the conventional Seebeck signal of
a Pt layer but should be detectable with sensitive techniques. In
addition, the skyrmion will acquire transverse Hall-like motion $v_{y}\approx-0.001\mbox{m/s}$
resulting in the Nernst response given by $V_{\text{ISH}}^{\perp}/W=V_{\text{ISH}}\sin\theta_{H}/W=2\times10^{-4}\mbox{V}/\mbox{m}$
which can be measured in the setup in Fig.~\ref{fig:Fig1}b).

We repeat the same estimate for $\text{MnSi}$ for which we take $a=0.5\mbox{nm}$,
$\mathfrak{s}=0.4\hbar/a^{3}$, $A/(a^{2}k_{B})=12\mbox{K}$, $Da=0.18A$,
$\xi\approx18\mbox{nm}$ \cite{Muhlbauer:Science2009}, $T=30\mbox{K}$
and a gradient $\partial_{\alpha}T=1\mbox{K}/\mu\mbox{m}$ arriving
at the skyrmion longitudinal velocity $v_{x}=0.02\mbox{m/s}$ in the
absence of thermoelectric spin currents. The thermoelectric spin current
$\mathbf{j}_{e}^{s}$ arising under the open circuit conditions in
conducting materials (i.e. in the absence of charge currents) \cite{Kovalev:PRB2009}
leads to additional contribution to velocity in Eqs. (\ref{eq:velocityX})
and (\ref{eq:velocityY}). Taking the Seebeck coefficient $S=10\mu\mbox{V}/\mbox{K}$,
its polarization $\wp_{S}=0.3$ and the conductivity $\sigma_{\text{MnSi}}=3\mu\Omega^{-1}\mbox{m}^{-1}$
we arrive at $v_{x}=-0.01\mbox{m/s}$ which shows that this contribution
pushes the domain wall away from the hot region when $\wp_{S}S>0$.
Since thermomagnonic forces strongly depend on the temperature it
seems to be feasible to reverse the direction of longitudinal skyrmion
motion at lower temperatures. For estimating the transverse motion
we use $\beta_{e}\approx\alpha$ for which the magnonic contribution
dominates in Eq.~(\ref{eq:velocityY}) arriving at $v_{y}\approx-0.002\mbox{m/s}$
and the Nernst response $V_{\text{ISH}}^{\perp}/W=V_{\text{ISH}}\sin\theta_{H}/W=10^{-3}\mbox{V}/\mbox{m}$. 

\textit{Conclusions.} We derived thermomagnonic torque and its
``$\beta$-type'' dissipative correction from the stochastic LLG
equation. Such corrections are important for magnetic
texture dynamics and here we show that they influence the sign of
the magnus force acting on skyrmion under temperature gradient. For
experimental detection we propose to use the transverse spin Seebeck
effect geometry (see Fig.~\ref{fig:Fig1}). Parametric excitation of spin waves of sufficiently short wavelength can also be described by our theory while effects related to linear momentum transfer should be included for longer wavelength \cite{Yan.Kamra.ea:PRB2013}. Our theory provides the
minimalistic phenomenological description while further studies could
incorporate magnon-magnon, magnon-phonon and magnon scattering on
impurities by constructing a microscopic kinetic theory. 

We acknowledge discussions with G. E. W. Bauer, Y. Tserkovnyak, O. Tretiakov,
and K. Belashchenko. In this version, Eqs. (\ref{eq:torque}), (\ref{eq:spin-accumulation}), and (\ref{eq:spin-accumulation1}) have been corrected following the footnote comment in Ref. \cite{Kim.Tserkovnyak:apa2015}. The research was performed in part
at the Central Facilities of the Nebraska Center for Materials and
Nanoscience supported by the Nebraska Research Initiative.

\bibliographystyle{apsrev}
\bibliography{Berryheat,MyBIB}

\begin{thebibliography}{50}
\expandafter\ifx\csname natexlab\endcsname\relax\def\natexlab#1{#1}\fi
\expandafter\ifx\csname bibnamefont\endcsname\relax
  \def\bibnamefont#1{#1}\fi
\expandafter\ifx\csname bibfnamefont\endcsname\relax
  \def\bibfnamefont#1{#1}\fi
\expandafter\ifx\csname citenamefont\endcsname\relax
  \def\citenamefont#1{#1}\fi
\expandafter\ifx\csname url\endcsname\relax
  \def\url#1{\texttt{#1}}\fi
\expandafter\ifx\csname urlprefix\endcsname\relax\def\urlprefix{URL }\fi
\providecommand{\bibinfo}[2]{#2}
\providecommand{\eprint}[2][]{\url{#2}}

\bibitem[{\citenamefont{Slonczewski}(1996)}]{Slonczewski:JMMM1996}
\bibinfo{author}{\bibfnamefont{J.}~\bibnamefont{Slonczewski}},
  \bibinfo{journal}{J. Magn. Magn. Mater.} \textbf{\bibinfo{volume}{159}},
  \bibinfo{pages}{L1 } (\bibinfo{year}{1996}).

\bibitem[{\citenamefont{Berger}(1996)}]{Berger:Oct1996}
\bibinfo{author}{\bibfnamefont{L.}~\bibnamefont{Berger}},
  \bibinfo{journal}{Phys. Rev. B} \textbf{\bibinfo{volume}{54}},
  \bibinfo{pages}{9353} (\bibinfo{year}{1996}).

\bibitem[{\citenamefont{Tsoi et~al.}(1998)\citenamefont{Tsoi, Jansen, Bass,
  Chiang, Seck, Tsoi, and Wyder}}]{Tsoi.Jansen.ea:PRL1998}
\bibinfo{author}{\bibfnamefont{M.}~\bibnamefont{Tsoi}},
  \bibinfo{author}{\bibfnamefont{A.}~\bibnamefont{Jansen}},
  \bibinfo{author}{\bibfnamefont{J.}~\bibnamefont{Bass}},
  \bibinfo{author}{\bibfnamefont{W.-C.} \bibnamefont{Chiang}},
  \bibinfo{author}{\bibfnamefont{M.}~\bibnamefont{Seck}},
  \bibinfo{author}{\bibfnamefont{V.}~\bibnamefont{Tsoi}}, \bibnamefont{and}
  \bibinfo{author}{\bibfnamefont{P.}~\bibnamefont{Wyder}},
  \bibinfo{journal}{Phys. Rev. Lett.} \textbf{\bibinfo{volume}{80}},
  \bibinfo{pages}{4281} (\bibinfo{year}{1998}).

\bibitem[{\citenamefont{Myers}(1999)}]{Myers:S1999}
\bibinfo{author}{\bibfnamefont{E.~B.} \bibnamefont{Myers}},
  \bibinfo{journal}{Science} \textbf{\bibinfo{volume}{285}},
  \bibinfo{pages}{867} (\bibinfo{year}{1999}).

\bibitem[{\citenamefont{{Tserkovnyak} et~al.}(2002)\citenamefont{{Tserkovnyak},
  {Brataas}, and {Bauer}}}]{Tserkovnyak:PRL2002}
\bibinfo{author}{\bibfnamefont{Y.}~\bibnamefont{{Tserkovnyak}}},
  \bibinfo{author}{\bibfnamefont{A.}~\bibnamefont{{Brataas}}},
  \bibnamefont{and} \bibinfo{author}{\bibfnamefont{G.~E.}
  \bibnamefont{{Bauer}}}, \bibinfo{journal}{Phys. Rev. Lett.}
  \textbf{\bibinfo{volume}{88}}, \bibinfo{eid}{117601} (\bibinfo{year}{2002}).

\bibitem[{\citenamefont{Heinrich et~al.}(2003)\citenamefont{Heinrich,
  Tserkovnyak, Woltersdorf, Brataas, Urban, and Bauer}}]{Heinrich:May2003}
\bibinfo{author}{\bibfnamefont{B.}~\bibnamefont{Heinrich}},
  \bibinfo{author}{\bibfnamefont{Y.}~\bibnamefont{Tserkovnyak}},
  \bibinfo{author}{\bibfnamefont{G.}~\bibnamefont{Woltersdorf}},
  \bibinfo{author}{\bibfnamefont{A.}~\bibnamefont{Brataas}},
  \bibinfo{author}{\bibfnamefont{R.}~\bibnamefont{Urban}}, \bibnamefont{and}
  \bibinfo{author}{\bibfnamefont{G.~E.~W.} \bibnamefont{Bauer}},
  \bibinfo{journal}{Phys. Rev. Lett.} \textbf{\bibinfo{volume}{90}},
  \bibinfo{pages}{187601} (\bibinfo{year}{2003}).

\bibitem[{\citenamefont{{Uchida} et~al.}(2008)\citenamefont{{Uchida},
  {Takahashi}, {Harii}, {Ieda}, {Koshibae}, {Ando}, {Maekawa}, and
  {Saitoh}}}]{Uchida.Takahashi.ea:2008}
\bibinfo{author}{\bibfnamefont{K.}~\bibnamefont{{Uchida}}},
  \bibinfo{author}{\bibfnamefont{S.}~\bibnamefont{{Takahashi}}},
  \bibinfo{author}{\bibfnamefont{K.}~\bibnamefont{{Harii}}},
  \bibinfo{author}{\bibfnamefont{J.}~\bibnamefont{{Ieda}}},
  \bibinfo{author}{\bibfnamefont{W.}~\bibnamefont{{Koshibae}}},
  \bibinfo{author}{\bibfnamefont{K.}~\bibnamefont{{Ando}}},
  \bibinfo{author}{\bibfnamefont{S.}~\bibnamefont{{Maekawa}}},
  \bibnamefont{and} \bibinfo{author}{\bibfnamefont{E.}~\bibnamefont{{Saitoh}}},
  \bibinfo{journal}{Nature} \textbf{\bibinfo{volume}{455}},
  \bibinfo{pages}{778} (\bibinfo{year}{2008}).

\bibitem[{\citenamefont{{Uchida} et~al.}(2010)\citenamefont{{Uchida}, {Ota},
  {Harii}, {Ando}, {Nakayama}, and {Saitoh}}}]{Uchida.Ota.ea:JoAP2010}
\bibinfo{author}{\bibfnamefont{K.}~\bibnamefont{{Uchida}}},
  \bibinfo{author}{\bibfnamefont{T.}~\bibnamefont{{Ota}}},
  \bibinfo{author}{\bibfnamefont{K.}~\bibnamefont{{Harii}}},
  \bibinfo{author}{\bibfnamefont{K.}~\bibnamefont{{Ando}}},
  \bibinfo{author}{\bibfnamefont{H.}~\bibnamefont{{Nakayama}}},
  \bibnamefont{and} \bibinfo{author}{\bibfnamefont{E.}~\bibnamefont{{Saitoh}}},
  \bibinfo{journal}{J. Appl. Phys.} \textbf{\bibinfo{volume}{107}},
  \bibinfo{pages}{090000} (\bibinfo{year}{2010}).

\bibitem[{\citenamefont{{Jaworski} et~al.}(2010)\citenamefont{{Jaworski},
  {Yang}, {Mack}, {Awschalom}, {Heremans}, and
  {Myers}}}]{Jaworski.Yang.ea:NM2010}
\bibinfo{author}{\bibfnamefont{C.~M.} \bibnamefont{{Jaworski}}},
  \bibinfo{author}{\bibfnamefont{J.}~\bibnamefont{{Yang}}},
  \bibinfo{author}{\bibfnamefont{S.}~\bibnamefont{{Mack}}},
  \bibinfo{author}{\bibfnamefont{D.~D.} \bibnamefont{{Awschalom}}},
  \bibinfo{author}{\bibfnamefont{J.~P.} \bibnamefont{{Heremans}}},
  \bibnamefont{and} \bibinfo{author}{\bibfnamefont{R.~C.}
  \bibnamefont{{Myers}}}, \bibinfo{journal}{Nat. Mater.}
  \textbf{\bibinfo{volume}{9}}, \bibinfo{pages}{898} (\bibinfo{year}{2010}).

\bibitem[{\citenamefont{Uchida et~al.}(2010)\citenamefont{Uchida, Xiao, Adachi,
  Ohe, Takahashi, Ieda, Ota, Kajiwara, Umezawa, Kawai et~al.}}]{Uchida:nov2010}
\bibinfo{author}{\bibfnamefont{K.}~\bibnamefont{Uchida}},
  \bibinfo{author}{\bibfnamefont{J.}~\bibnamefont{Xiao}},
  \bibinfo{author}{\bibfnamefont{H.}~\bibnamefont{Adachi}},
  \bibinfo{author}{\bibfnamefont{J.}~\bibnamefont{Ohe}},
  \bibinfo{author}{\bibfnamefont{S.}~\bibnamefont{Takahashi}},
  \bibinfo{author}{\bibfnamefont{J.}~\bibnamefont{Ieda}},
  \bibinfo{author}{\bibfnamefont{T.}~\bibnamefont{Ota}},
  \bibinfo{author}{\bibfnamefont{Y.}~\bibnamefont{Kajiwara}},
  \bibinfo{author}{\bibfnamefont{H.}~\bibnamefont{Umezawa}},
  \bibinfo{author}{\bibfnamefont{H.}~\bibnamefont{Kawai}},
  \bibnamefont{et~al.}, \bibinfo{journal}{Nat Mater}
  \textbf{\bibinfo{volume}{9}}, \bibinfo{pages}{894} (\bibinfo{year}{2010}).

\bibitem[{\citenamefont{{Bosu} et~al.}(2011)\citenamefont{{Bosu}, {Sakuraba},
  {Uchida}, {Saito}, {Ota}, {Saitoh}, and {Takanashi}}}]{Bosu.Sakuraba.ea:2011}
\bibinfo{author}{\bibfnamefont{S.}~\bibnamefont{{Bosu}}},
  \bibinfo{author}{\bibfnamefont{Y.}~\bibnamefont{{Sakuraba}}},
  \bibinfo{author}{\bibfnamefont{K.}~\bibnamefont{{Uchida}}},
  \bibinfo{author}{\bibfnamefont{K.}~\bibnamefont{{Saito}}},
  \bibinfo{author}{\bibfnamefont{T.}~\bibnamefont{{Ota}}},
  \bibinfo{author}{\bibfnamefont{E.}~\bibnamefont{{Saitoh}}}, \bibnamefont{and}
  \bibinfo{author}{\bibfnamefont{K.}~\bibnamefont{{Takanashi}}},
  \bibinfo{journal}{Phys. Rev. B} \textbf{\bibinfo{volume}{83}},
  \bibinfo{eid}{224401} (\bibinfo{year}{2011}).

\bibitem[{\citenamefont{Maekawa et~al.}(2013)\citenamefont{Maekawa, Adachi,
  Uchida, Ieda, and Saitoh}}]{Maekawa.Adachi.ea:JPSJ2013}
\bibinfo{author}{\bibfnamefont{S.}~\bibnamefont{Maekawa}},
  \bibinfo{author}{\bibfnamefont{H.}~\bibnamefont{Adachi}},
  \bibinfo{author}{\bibfnamefont{K.-I.} \bibnamefont{Uchida}},
  \bibinfo{author}{\bibfnamefont{J.}~\bibnamefont{Ieda}}, \bibnamefont{and}
  \bibinfo{author}{\bibfnamefont{E.}~\bibnamefont{Saitoh}},
  \bibinfo{journal}{J. Phys. Soc. Jpn.} \textbf{\bibinfo{volume}{82}},
  \bibinfo{pages}{102002} (\bibinfo{year}{2013}).

\bibitem[{\citenamefont{Parkin et~al.}(2008)\citenamefont{Parkin, Hayashi, and
  Thomas}}]{Parkin.Hayashi.ea:S2008}
\bibinfo{author}{\bibfnamefont{S.~S.~P.} \bibnamefont{Parkin}},
  \bibinfo{author}{\bibfnamefont{M.}~\bibnamefont{Hayashi}}, \bibnamefont{and}
  \bibinfo{author}{\bibfnamefont{L.}~\bibnamefont{Thomas}},
  \bibinfo{journal}{Science} \textbf{\bibinfo{volume}{320}},
  \bibinfo{pages}{190} (\bibinfo{year}{2008}).

\bibitem[{\citenamefont{Bauer et~al.}(2012)\citenamefont{Bauer, Saitoh, and van
  Wees}}]{Bauer.Saitoh.ea:NM2012}
\bibinfo{author}{\bibfnamefont{G.~E.~W.} \bibnamefont{Bauer}},
  \bibinfo{author}{\bibfnamefont{E.}~\bibnamefont{Saitoh}}, \bibnamefont{and}
  \bibinfo{author}{\bibfnamefont{B.~J.} \bibnamefont{van Wees}},
  \bibinfo{journal}{Nat. Mater.} \textbf{\bibinfo{volume}{11}},
  \bibinfo{pages}{391} (\bibinfo{year}{2012}).

\bibitem[{\citenamefont{Kovalev and Tserkovnyak}(2009)}]{Kovalev:PRB2009}
\bibinfo{author}{\bibfnamefont{A.~A.} \bibnamefont{Kovalev}} \bibnamefont{and}
  \bibinfo{author}{\bibfnamefont{Y.}~\bibnamefont{Tserkovnyak}},
  \bibinfo{journal}{Phys. Rev. B} \textbf{\bibinfo{volume}{80}},
  \bibinfo{pages}{100408} (\bibinfo{year}{2009}).

\bibitem[{\citenamefont{Bauer et~al.}(2010)\citenamefont{Bauer, Bretzel,
  Brataas, and Tserkovnyak}}]{Bauer:jan2010}
\bibinfo{author}{\bibfnamefont{G.~E.~W.} \bibnamefont{Bauer}},
  \bibinfo{author}{\bibfnamefont{S.}~\bibnamefont{Bretzel}},
  \bibinfo{author}{\bibfnamefont{A.}~\bibnamefont{Brataas}}, \bibnamefont{and}
  \bibinfo{author}{\bibfnamefont{Y.}~\bibnamefont{Tserkovnyak}},
  \bibinfo{journal}{Phys. Rev. B} \textbf{\bibinfo{volume}{81}},
  \bibinfo{pages}{024427} (\bibinfo{year}{2010}).

\bibitem[{\citenamefont{Hinzke and Nowak}(2011)}]{Hinzke:PRL2011}
\bibinfo{author}{\bibfnamefont{D.}~\bibnamefont{Hinzke}} \bibnamefont{and}
  \bibinfo{author}{\bibfnamefont{U.}~\bibnamefont{Nowak}},
  \bibinfo{journal}{Phys. Rev. Lett.} \textbf{\bibinfo{volume}{107}},
  \bibinfo{pages}{027205} (\bibinfo{year}{2011}).

\bibitem[{\citenamefont{Kovalev and Tserkovnyak}(2012)}]{Kovalev:EPL2012}
\bibinfo{author}{\bibfnamefont{A.~A.} \bibnamefont{Kovalev}} \bibnamefont{and}
  \bibinfo{author}{\bibfnamefont{Y.}~\bibnamefont{Tserkovnyak}},
  \bibinfo{journal}{EPL (Europhysics Letters)} \textbf{\bibinfo{volume}{97}},
  \bibinfo{pages}{67002} (\bibinfo{year}{2012}).

\bibitem[{\citenamefont{Torrejon et~al.}(2012)\citenamefont{Torrejon,
  Malinowski, Pelloux, Weil, Thiaville, Curiale, Lacour, Montaigne, and
  Hehn}}]{Torrejon.Malinowski.ea:PRL2012}
\bibinfo{author}{\bibfnamefont{J.}~\bibnamefont{Torrejon}},
  \bibinfo{author}{\bibfnamefont{G.}~\bibnamefont{Malinowski}},
  \bibinfo{author}{\bibfnamefont{M.}~\bibnamefont{Pelloux}},
  \bibinfo{author}{\bibfnamefont{R.}~\bibnamefont{Weil}},
  \bibinfo{author}{\bibfnamefont{A.}~\bibnamefont{Thiaville}},
  \bibinfo{author}{\bibfnamefont{J.}~\bibnamefont{Curiale}},
  \bibinfo{author}{\bibfnamefont{D.}~\bibnamefont{Lacour}},
  \bibinfo{author}{\bibfnamefont{F.}~\bibnamefont{Montaigne}},
  \bibnamefont{and} \bibinfo{author}{\bibfnamefont{M.}~\bibnamefont{Hehn}},
  \bibinfo{journal}{Phys. Rev. Lett.} \textbf{\bibinfo{volume}{109}}
  (\bibinfo{year}{2012}).

\bibitem[{\citenamefont{Jiang et~al.}(2013)\citenamefont{Jiang, Upadhyaya, Fan,
  Zhao, Wang, Chang, Lang, Wong, Lewis, Lin et~al.}}]{Jiang:PRL2013}
\bibinfo{author}{\bibfnamefont{W.}~\bibnamefont{Jiang}},
  \bibinfo{author}{\bibfnamefont{P.}~\bibnamefont{Upadhyaya}},
  \bibinfo{author}{\bibfnamefont{Y.}~\bibnamefont{Fan}},
  \bibinfo{author}{\bibfnamefont{J.}~\bibnamefont{Zhao}},
  \bibinfo{author}{\bibfnamefont{M.}~\bibnamefont{Wang}},
  \bibinfo{author}{\bibfnamefont{L.-T.} \bibnamefont{Chang}},
  \bibinfo{author}{\bibfnamefont{M.}~\bibnamefont{Lang}},
  \bibinfo{author}{\bibfnamefont{K.~L.} \bibnamefont{Wong}},
  \bibinfo{author}{\bibfnamefont{M.}~\bibnamefont{Lewis}},
  \bibinfo{author}{\bibfnamefont{Y.-T.} \bibnamefont{Lin}},
  \bibnamefont{et~al.}, \bibinfo{journal}{Phys. Rev. Lett.}
  \textbf{\bibinfo{volume}{110}}, \bibinfo{pages}{177202}
  (\bibinfo{year}{2013}).

\bibitem[{\citenamefont{Cahaya et~al.}(2014)\citenamefont{Cahaya, Tretiakov,
  and Bauer}}]{Cahaya.Tretiakov.ea:APL2014}
\bibinfo{author}{\bibfnamefont{A.~B.} \bibnamefont{Cahaya}},
  \bibinfo{author}{\bibfnamefont{O.~A.} \bibnamefont{Tretiakov}},
  \bibnamefont{and} \bibinfo{author}{\bibfnamefont{G.~E.~W.}
  \bibnamefont{Bauer}}, \bibinfo{journal}{Appl. Phys. Lett.}
  \textbf{\bibinfo{volume}{104}}, \bibinfo{pages}{042402}
  (\bibinfo{year}{2014}).

\bibitem[{\citenamefont{Kovalev and Tserkovnyak}(2010)}]{Kovalev:SSC2010}
\bibinfo{author}{\bibfnamefont{A.~A.} \bibnamefont{Kovalev}} \bibnamefont{and}
  \bibinfo{author}{\bibfnamefont{Y.}~\bibnamefont{Tserkovnyak}},
  \bibinfo{journal}{Solid State Commun.} \textbf{\bibinfo{volume}{150}},
  \bibinfo{pages}{500} (\bibinfo{year}{2010}).

\bibitem[{\citenamefont{Seki et~al.}(2012)\citenamefont{Seki, Yu, Ishiwata, and
  Tokura}}]{Seki:Science2012}
\bibinfo{author}{\bibfnamefont{S.}~\bibnamefont{Seki}},
  \bibinfo{author}{\bibfnamefont{X.~Z.} \bibnamefont{Yu}},
  \bibinfo{author}{\bibfnamefont{S.}~\bibnamefont{Ishiwata}}, \bibnamefont{and}
  \bibinfo{author}{\bibfnamefont{Y.}~\bibnamefont{Tokura}},
  \bibinfo{journal}{Science} \textbf{\bibinfo{volume}{336}},
  \bibinfo{pages}{198} (\bibinfo{year}{2012}).

\bibitem[{\citenamefont{{Yu} et~al.}(2012)\citenamefont{{Yu}, {Mostovoy},
  {Tokunaga}, {Zhang}, {Kimoto}, {Matsui}, {Kaneko}, {Nagaosa}, and
  {Tokura}}}]{Yu.Mostovoy.ea:PotNAoS2012}
\bibinfo{author}{\bibfnamefont{X.}~\bibnamefont{{Yu}}},
  \bibinfo{author}{\bibfnamefont{M.}~\bibnamefont{{Mostovoy}}},
  \bibinfo{author}{\bibfnamefont{Y.}~\bibnamefont{{Tokunaga}}},
  \bibinfo{author}{\bibfnamefont{W.}~\bibnamefont{{Zhang}}},
  \bibinfo{author}{\bibfnamefont{K.}~\bibnamefont{{Kimoto}}},
  \bibinfo{author}{\bibfnamefont{Y.}~\bibnamefont{{Matsui}}},
  \bibinfo{author}{\bibfnamefont{Y.}~\bibnamefont{{Kaneko}}},
  \bibinfo{author}{\bibfnamefont{N.}~\bibnamefont{{Nagaosa}}},
  \bibnamefont{and} \bibinfo{author}{\bibfnamefont{Y.}~\bibnamefont{{Tokura}}},
  \bibinfo{journal}{Proc. Natl. Acad. Sci. USA} \textbf{\bibinfo{volume}{109}},
  \bibinfo{pages}{8856} (\bibinfo{year}{2012}).

\bibitem[{\citenamefont{Bhagat et~al.}(1973)\citenamefont{Bhagat, Lessoff,
  Vittoria, and Guenzer}}]{Bhagat.Lessoff.ea:PSSa1973}
\bibinfo{author}{\bibfnamefont{S.}~\bibnamefont{Bhagat}},
  \bibinfo{author}{\bibfnamefont{H.}~\bibnamefont{Lessoff}},
  \bibinfo{author}{\bibfnamefont{C.}~\bibnamefont{Vittoria}}, \bibnamefont{and}
  \bibinfo{author}{\bibfnamefont{C.}~\bibnamefont{Guenzer}},
  \bibinfo{journal}{Phys. Stat. Sol. (a)} \textbf{\bibinfo{volume}{20}},
  \bibinfo{pages}{731} (\bibinfo{year}{1973}).

\bibitem[{\citenamefont{Hoffman et~al.}(2013)\citenamefont{Hoffman, Sato, and
  Tserkovnyak}}]{Hoffman.Sato.ea:PRB2013}
\bibinfo{author}{\bibfnamefont{S.}~\bibnamefont{Hoffman}},
  \bibinfo{author}{\bibfnamefont{K.}~\bibnamefont{Sato}}, \bibnamefont{and}
  \bibinfo{author}{\bibfnamefont{Y.}~\bibnamefont{Tserkovnyak}},
  \bibinfo{journal}{Phys. Rev. B} \textbf{\bibinfo{volume}{88}},
  \bibinfo{pages}{064408} (\bibinfo{year}{2013}).

\bibitem[{\citenamefont{Dugaev et~al.}(2005)\citenamefont{Dugaev, Bruno,
  Canals, and Lacroix}}]{Dugaev:jul2005}
\bibinfo{author}{\bibfnamefont{V.~K.} \bibnamefont{Dugaev}},
  \bibinfo{author}{\bibfnamefont{P.}~\bibnamefont{Bruno}},
  \bibinfo{author}{\bibfnamefont{B.}~\bibnamefont{Canals}}, \bibnamefont{and}
  \bibinfo{author}{\bibfnamefont{C.}~\bibnamefont{Lacroix}},
  \bibinfo{journal}{Phys. Rev. B} \textbf{\bibinfo{volume}{72}},
  \bibinfo{pages}{024456} (\bibinfo{year}{2005}).

\bibitem[{\citenamefont{Zhang and Li}(2004)}]{Zhang.Li:PRL2004}
\bibinfo{author}{\bibfnamefont{S.}~\bibnamefont{Zhang}} \bibnamefont{and}
  \bibinfo{author}{\bibfnamefont{Z.}~\bibnamefont{Li}}, \bibinfo{journal}{Phys.
  Rev. Lett.} \textbf{\bibinfo{volume}{93}}, \bibinfo{pages}{127204}
  (\bibinfo{year}{2004}).

\bibitem[{\citenamefont{Kohno et~al.}(2006)\citenamefont{Kohno, Tatara, and
  Shibata}}]{Kohno.Tatara.ea:JotPSoJ2006}
\bibinfo{author}{\bibfnamefont{H.}~\bibnamefont{Kohno}},
  \bibinfo{author}{\bibfnamefont{G.}~\bibnamefont{Tatara}}, \bibnamefont{and}
  \bibinfo{author}{\bibfnamefont{J.}~\bibnamefont{Shibata}},
  \bibinfo{journal}{J. Phys. Soc. Jpn.} \textbf{\bibinfo{volume}{75}},
  \bibinfo{pages}{113706} (\bibinfo{year}{2006}).

\bibitem[{\citenamefont{Tserkovnyak et~al.}(2006)\citenamefont{Tserkovnyak,
  Skadsem, Brataas, and Bauer}}]{Tserkovnyak.Skadsem.ea:PRB2006}
\bibinfo{author}{\bibfnamefont{Y.}~\bibnamefont{Tserkovnyak}},
  \bibinfo{author}{\bibfnamefont{H.}~\bibnamefont{Skadsem}},
  \bibinfo{author}{\bibfnamefont{A.}~\bibnamefont{Brataas}}, \bibnamefont{and}
  \bibinfo{author}{\bibfnamefont{G.}~\bibnamefont{Bauer}},
  \bibinfo{journal}{Phys. Rev. B} \textbf{\bibinfo{volume}{74}}
  (\bibinfo{year}{2006}).

\bibitem[{\citenamefont{Tserkovnyak et~al.}(2008)\citenamefont{Tserkovnyak,
  Brataas, and Bauer}}]{Tserkovnyak.Brataas.ea:JoMaMM2008}
\bibinfo{author}{\bibfnamefont{Y.}~\bibnamefont{Tserkovnyak}},
  \bibinfo{author}{\bibfnamefont{A.}~\bibnamefont{Brataas}}, \bibnamefont{and}
  \bibinfo{author}{\bibfnamefont{G.~E.} \bibnamefont{Bauer}},
  \bibinfo{journal}{J. Magn. Magn. Mater.} \textbf{\bibinfo{volume}{320}},
  \bibinfo{pages}{1282} (\bibinfo{year}{2008}).

\bibitem[{\citenamefont{Hals et~al.}(2010)\citenamefont{Hals, Brataas, and
  Bauer}}]{Hals.Brataas.ea:SSC2010}
\bibinfo{author}{\bibfnamefont{K.~M.} \bibnamefont{Hals}},
  \bibinfo{author}{\bibfnamefont{A.}~\bibnamefont{Brataas}}, \bibnamefont{and}
  \bibinfo{author}{\bibfnamefont{G.~E.} \bibnamefont{Bauer}},
  \bibinfo{journal}{Solid State Commun.} \textbf{\bibinfo{volume}{150}},
  \bibinfo{pages}{461} (\bibinfo{year}{2010}).

\bibitem[{\citenamefont{{R{\"o}{\ss}ler}
  et~al.}(2006)\citenamefont{{R{\"o}{\ss}ler}, {Bogdanov}, and
  {Pfleiderer}}}]{Rossler:2006}
\bibinfo{author}{\bibfnamefont{U.~K.} \bibnamefont{{R{\"o}{\ss}ler}}},
  \bibinfo{author}{\bibfnamefont{A.~N.} \bibnamefont{{Bogdanov}}},
  \bibnamefont{and}
  \bibinfo{author}{\bibfnamefont{C.}~\bibnamefont{{Pfleiderer}}},
  \bibinfo{journal}{Nature} \textbf{\bibinfo{volume}{442}},
  \bibinfo{pages}{797} (\bibinfo{year}{2006}).

\bibitem[{\citenamefont{{M{\"u}hlbauer}
  et~al.}(2009)\citenamefont{{M{\"u}hlbauer}, {Binz}, {Jonietz}, {Pfleiderer},
  {Rosch}, {Neubauer}, {Georgii}, and {B{\"o}ni}}}]{Muhlbauer:Science2009}
\bibinfo{author}{\bibfnamefont{S.}~\bibnamefont{{M{\"u}hlbauer}}},
  \bibinfo{author}{\bibfnamefont{B.}~\bibnamefont{{Binz}}},
  \bibinfo{author}{\bibfnamefont{F.}~\bibnamefont{{Jonietz}}},
  \bibinfo{author}{\bibfnamefont{C.}~\bibnamefont{{Pfleiderer}}},
  \bibinfo{author}{\bibfnamefont{A.}~\bibnamefont{{Rosch}}},
  \bibinfo{author}{\bibfnamefont{A.}~\bibnamefont{{Neubauer}}},
  \bibinfo{author}{\bibfnamefont{R.}~\bibnamefont{{Georgii}}},
  \bibnamefont{and}
  \bibinfo{author}{\bibfnamefont{P.}~\bibnamefont{{B{\"o}ni}}},
  \bibinfo{journal}{Science} \textbf{\bibinfo{volume}{323}},
  \bibinfo{pages}{915} (\bibinfo{year}{2009}).

\bibitem[{\citenamefont{{Yi} et~al.}(2009)\citenamefont{{Yi}, {Onoda},
  {Nagaosa}, and {Han}}}]{Yi:2009}
\bibinfo{author}{\bibfnamefont{S.~D.} \bibnamefont{{Yi}}},
  \bibinfo{author}{\bibfnamefont{S.}~\bibnamefont{{Onoda}}},
  \bibinfo{author}{\bibfnamefont{N.}~\bibnamefont{{Nagaosa}}},
  \bibnamefont{and} \bibinfo{author}{\bibfnamefont{J.~H.} \bibnamefont{{Han}}},
  \bibinfo{journal}{Phys. Rev. B} \textbf{\bibinfo{volume}{80}},
  \bibinfo{eid}{054416} (\bibinfo{year}{2009}).

\bibitem[{\citenamefont{{Butenko} et~al.}(2010)\citenamefont{{Butenko},
  {Leonov}, {R{\"o}{\ss}ler}, and {Bogdanov}}}]{Butenko:2010}
\bibinfo{author}{\bibfnamefont{A.~B.} \bibnamefont{{Butenko}}},
  \bibinfo{author}{\bibfnamefont{A.~A.} \bibnamefont{{Leonov}}},
  \bibinfo{author}{\bibfnamefont{U.~K.} \bibnamefont{{R{\"o}{\ss}ler}}},
  \bibnamefont{and} \bibinfo{author}{\bibfnamefont{A.~N.}
  \bibnamefont{{Bogdanov}}}, \bibinfo{journal}{Phys. Rev. B}
  \textbf{\bibinfo{volume}{82}}, \bibinfo{eid}{052403} (\bibinfo{year}{2010}).

\bibitem[{\citenamefont{{Yu} et~al.}(2010)\citenamefont{{Yu}, {Onose},
  {Kanazawa}, {Park}, {Han}, {Matsui}, {Nagaosa}, and
  {Tokura}}}]{Yu:Nature2010}
\bibinfo{author}{\bibfnamefont{X.~Z.} \bibnamefont{{Yu}}},
  \bibinfo{author}{\bibfnamefont{Y.}~\bibnamefont{{Onose}}},
  \bibinfo{author}{\bibfnamefont{N.}~\bibnamefont{{Kanazawa}}},
  \bibinfo{author}{\bibfnamefont{J.~H.} \bibnamefont{{Park}}},
  \bibinfo{author}{\bibfnamefont{J.~H.} \bibnamefont{{Han}}},
  \bibinfo{author}{\bibfnamefont{Y.}~\bibnamefont{{Matsui}}},
  \bibinfo{author}{\bibfnamefont{N.}~\bibnamefont{{Nagaosa}}},
  \bibnamefont{and} \bibinfo{author}{\bibfnamefont{Y.}~\bibnamefont{{Tokura}}},
  \bibinfo{journal}{Nature} \textbf{\bibinfo{volume}{465}},
  \bibinfo{pages}{901} (\bibinfo{year}{2010}).

\bibitem[{\citenamefont{{Yu} et~al.}(2011)\citenamefont{{Yu}, {Kanazawa},
  {Onose}, {Kimoto}, {Zhang}, {Ishiwata}, {Matsui}, and {Tokura}}}]{Yu:NM2011}
\bibinfo{author}{\bibfnamefont{X.~Z.} \bibnamefont{{Yu}}},
  \bibinfo{author}{\bibfnamefont{N.}~\bibnamefont{{Kanazawa}}},
  \bibinfo{author}{\bibfnamefont{Y.}~\bibnamefont{{Onose}}},
  \bibinfo{author}{\bibfnamefont{K.}~\bibnamefont{{Kimoto}}},
  \bibinfo{author}{\bibfnamefont{W.~Z.} \bibnamefont{{Zhang}}},
  \bibinfo{author}{\bibfnamefont{S.}~\bibnamefont{{Ishiwata}}},
  \bibinfo{author}{\bibfnamefont{Y.}~\bibnamefont{{Matsui}}}, \bibnamefont{and}
  \bibinfo{author}{\bibfnamefont{Y.}~\bibnamefont{{Tokura}}},
  \bibinfo{journal}{Nat. Mater.} \textbf{\bibinfo{volume}{10}},
  \bibinfo{pages}{106} (\bibinfo{year}{2011}).

\bibitem[{\citenamefont{Mochizuki et~al.}(2014)\citenamefont{Mochizuki, Yu,
  Seki, Kanazawa, Koshibae, Zang, Mostovoy, Tokura, and
  Nagaosa}}]{Mochizuki.Yu.ea:NM2014}
\bibinfo{author}{\bibfnamefont{M.}~\bibnamefont{Mochizuki}},
  \bibinfo{author}{\bibfnamefont{X.~Z.} \bibnamefont{Yu}},
  \bibinfo{author}{\bibfnamefont{S.}~\bibnamefont{Seki}},
  \bibinfo{author}{\bibfnamefont{N.}~\bibnamefont{Kanazawa}},
  \bibinfo{author}{\bibfnamefont{W.}~\bibnamefont{Koshibae}},
  \bibinfo{author}{\bibfnamefont{J.}~\bibnamefont{Zang}},
  \bibinfo{author}{\bibfnamefont{M.}~\bibnamefont{Mostovoy}},
  \bibinfo{author}{\bibfnamefont{Y.}~\bibnamefont{Tokura}}, \bibnamefont{and}
  \bibinfo{author}{\bibfnamefont{N.}~\bibnamefont{Nagaosa}},
  \bibinfo{journal}{Nat. Mater.} \textbf{\bibinfo{volume}{13}},
  \bibinfo{pages}{241} (\bibinfo{year}{2014}).

\bibitem[{\citenamefont{{Everschor} et~al.}(2012)\citenamefont{{Everschor},
  {Garst}, {Binz}, {Jonietz}, {M{\"u}hlbauer}, {Pfleiderer}, and
  {Rosch}}}]{Everschor2012}
\bibinfo{author}{\bibfnamefont{K.}~\bibnamefont{{Everschor}}},
  \bibinfo{author}{\bibfnamefont{M.}~\bibnamefont{{Garst}}},
  \bibinfo{author}{\bibfnamefont{B.}~\bibnamefont{{Binz}}},
  \bibinfo{author}{\bibfnamefont{F.}~\bibnamefont{{Jonietz}}},
  \bibinfo{author}{\bibfnamefont{S.}~\bibnamefont{{M{\"u}hlbauer}}},
  \bibinfo{author}{\bibfnamefont{C.}~\bibnamefont{{Pfleiderer}}},
  \bibnamefont{and} \bibinfo{author}{\bibfnamefont{A.}~\bibnamefont{{Rosch}}},
  \bibinfo{journal}{Phys. Rev. B} \textbf{\bibinfo{volume}{86}},
  \bibinfo{eid}{054432} (\bibinfo{year}{2012}).

\bibitem[{\citenamefont{{Kong} and {Zang}}(2013)}]{Kong:PRL2013}
\bibinfo{author}{\bibfnamefont{L.}~\bibnamefont{{Kong}}} \bibnamefont{and}
  \bibinfo{author}{\bibfnamefont{J.}~\bibnamefont{{Zang}}},
  \bibinfo{journal}{Phys. Rev. Lett.} \textbf{\bibinfo{volume}{111}},
  \bibinfo{eid}{067203} (\bibinfo{year}{2013}).

\bibitem[{\citenamefont{Lin et~al.}(2014)\citenamefont{Lin, Batista,
  Reichhardt, and Saxena}}]{Saxena:arXiv2013}
\bibinfo{author}{\bibfnamefont{S.-Z.} \bibnamefont{Lin}},
  \bibinfo{author}{\bibfnamefont{C.~D.} \bibnamefont{Batista}},
  \bibinfo{author}{\bibfnamefont{C.}~\bibnamefont{Reichhardt}},
  \bibnamefont{and} \bibinfo{author}{\bibfnamefont{A.}~\bibnamefont{Saxena}},
  \bibinfo{journal}{Phys. Rev. Lett.} \textbf{\bibinfo{volume}{112}},
  \bibinfo{pages}{187203} (\bibinfo{year}{2014}).

\bibitem[{\citenamefont{{Brown}}(1963)}]{Brown:PR1963}
\bibinfo{author}{\bibfnamefont{W.~F.} \bibnamefont{{Brown}}},
  \bibinfo{journal}{Phys. Rev.} \textbf{\bibinfo{volume}{130}},
  \bibinfo{pages}{1677} (\bibinfo{year}{1963}).

\bibitem[{\citenamefont{Ashcroft and Mermin}(1976)}]{Ashcroft.Mermin:1976}
\bibinfo{author}{\bibfnamefont{N.~W.} \bibnamefont{Ashcroft}} \bibnamefont{and}
  \bibinfo{author}{\bibfnamefont{N.~D.} \bibnamefont{Mermin}},
  \emph{\bibinfo{title}{Solid State Physics}} (\bibinfo{publisher}{Cengage
  Learning}, \bibinfo{year}{1976}).

\bibitem[{\citenamefont{{Thiele}}(1973)}]{Thiele:PRL1973}
\bibinfo{author}{\bibfnamefont{A.~A.} \bibnamefont{{Thiele}}},
  \bibinfo{journal}{Phys. Rev. Lett.} \textbf{\bibinfo{volume}{30}},
  \bibinfo{pages}{230} (\bibinfo{year}{1973}).

\bibitem[{\citenamefont{{Xia} et~al.}(2002)\citenamefont{{Xia}, {Kelly},
  {Bauer}, {Brataas}, and {Turek}}}]{Xia.Kelly.ea:2002}
\bibinfo{author}{\bibfnamefont{K.}~\bibnamefont{{Xia}}},
  \bibinfo{author}{\bibfnamefont{P.~J.} \bibnamefont{{Kelly}}},
  \bibinfo{author}{\bibfnamefont{G.~E.} \bibnamefont{{Bauer}}},
  \bibinfo{author}{\bibfnamefont{A.}~\bibnamefont{{Brataas}}},
  \bibnamefont{and} \bibinfo{author}{\bibfnamefont{I.}~\bibnamefont{{Turek}}},
  \bibinfo{journal}{Phys. Rev. B} \textbf{\bibinfo{volume}{65}},
  \bibinfo{eid}{220401} (\bibinfo{year}{2002}).

\bibitem[{\citenamefont{Saitoh et~al.}(2006)\citenamefont{Saitoh, Ueda,
  Miyajima, and Tatara}}]{Saitoh:APL2006}
\bibinfo{author}{\bibfnamefont{E.}~\bibnamefont{Saitoh}},
  \bibinfo{author}{\bibfnamefont{M.}~\bibnamefont{Ueda}},
  \bibinfo{author}{\bibfnamefont{H.}~\bibnamefont{Miyajima}}, \bibnamefont{and}
  \bibinfo{author}{\bibfnamefont{G.}~\bibnamefont{Tatara}},
  \bibinfo{journal}{Appl. Phys. Lett.} \textbf{\bibinfo{volume}{88}},
  \bibinfo{eid}{182509} (\bibinfo{year}{2006}).

\bibitem[{\citenamefont{{Weiler} et~al.}(2013)\citenamefont{{Weiler},
  {Althammer}, {Schreier}, {Lotze}, {Pernpeintner}, {Meyer}, {Huebl}, {Gross},
  {Kamra}, {Xiao} et~al.}}]{Weiler:PRL2013}
\bibinfo{author}{\bibfnamefont{M.}~\bibnamefont{{Weiler}}},
  \bibinfo{author}{\bibfnamefont{M.}~\bibnamefont{{Althammer}}},
  \bibinfo{author}{\bibfnamefont{M.}~\bibnamefont{{Schreier}}},
  \bibinfo{author}{\bibfnamefont{J.}~\bibnamefont{{Lotze}}},
  \bibinfo{author}{\bibfnamefont{M.}~\bibnamefont{{Pernpeintner}}},
  \bibinfo{author}{\bibfnamefont{S.}~\bibnamefont{{Meyer}}},
  \bibinfo{author}{\bibfnamefont{H.}~\bibnamefont{{Huebl}}},
  \bibinfo{author}{\bibfnamefont{R.}~\bibnamefont{{Gross}}},
  \bibinfo{author}{\bibfnamefont{A.}~\bibnamefont{{Kamra}}},
  \bibinfo{author}{\bibfnamefont{J.}~\bibnamefont{{Xiao}}},
  \bibnamefont{et~al.}, \bibinfo{journal}{Phys. Rev. Lett.}
  \textbf{\bibinfo{volume}{111}}, \bibinfo{eid}{176601} (\bibinfo{year}{2013}),
  \eprint{1306.5012}.

\bibitem[{\citenamefont{Yan et~al.}(2013)\citenamefont{Yan, Kamra, Cao, and
  Bauer}}]{Yan.Kamra.ea:PRB2013}
\bibinfo{author}{\bibfnamefont{P.}~\bibnamefont{Yan}},
  \bibinfo{author}{\bibfnamefont{A.}~\bibnamefont{Kamra}},
  \bibinfo{author}{\bibfnamefont{Y.}~\bibnamefont{Cao}}, \bibnamefont{and}
  \bibinfo{author}{\bibfnamefont{G.~E.~W.} \bibnamefont{Bauer}},
  \bibinfo{journal}{Phys. Rev. B} \textbf{\bibinfo{volume}{88}},
  \bibinfo{pages}{144413} (\bibinfo{year}{2013}).

\bibitem[{\citenamefont{Kim and Tserkovnyak}(2015)}]{Kim.Tserkovnyak:apa2015}
\bibinfo{author}{\bibfnamefont{S.~K.} \bibnamefont{Kim}} \bibnamefont{and}
  \bibinfo{author}{\bibfnamefont{Y.}~\bibnamefont{Tserkovnyak}},
  \bibinfo{journal}{arXiv preprint arXiv:1505.00818}  (\bibinfo{year}{2015}).

\end{thebibliography}

\end{document}